\documentclass[graybox]{svmult}

\usepackage{mathptmx}       
\usepackage{helvet}         
\usepackage{courier}        
\usepackage{type1cm}        
\usepackage{makeidx}         
\usepackage{graphicx}        
\usepackage{multicol}        
\usepackage[bottom]{footmisc}

\newcommand\rf[1]{(\ref{eq:#1})}
\newcommand\lab[1]{\label{eq:#1}}
\newcommand\nonu{\nonumber}
\newcommand\br{\begin{eqnarray}}
\newcommand\er{\end{eqnarray}}
\newcommand\be{\begin{equation}}
\newcommand\ee{\end{equation}}

\newcommand\foot[1]{\footnotemark\footnotetext{#1}}
\newcommand\lb{\lbrack}
\newcommand\rb{\rbrack}

\newcommand\llb{\left\lbrack}
\newcommand\rrb{\right\rbrack}

\renewcommand\({\left(}
\renewcommand\){\right)}
\newcommand\bv{\bigm\vert}               

\newcommand\bc{\begin{center}}
\newcommand\ec{\end{center}}

\newcommand\partder[2]{\frac{{\partial {#1}}}{{\partial {#2}}}}

\renewcommand\a{\alpha}
\renewcommand\b{\beta}

\renewcommand\d{\delta}

\newcommand\g{\gamma}
\newcommand\G{\Gamma}

\newcommand\h{\frac{1}{2}}
\renewcommand\k{\kappa}
\renewcommand\l{\lambda}
\renewcommand\L{\Lambda}
\newcommand\m{\mu}
\newcommand\n{\nu}

\newcommand\vp{\varphi}
\renewcommand\P{\Phi}
\newcommand\pa{\partial}

\newcommand\pr{\prime}

\renewcommand\th{\theta}

\newcommand\wti{\widetilde}

\newcommand\cE{{\mathcal E}}

\newcommand\cM{{\mathcal M}}

\newcommand{\ct}[1]{\cite{#1}}
\newcommand{\bib}[1]{\bibitem{#1}}

\newcommand\PRL[3]{\textsl{Phys. Rev. Lett.} \textbf{#1} (#2) #3}
\newcommand\NPB[3]{\textsl{Nucl. Phys.} \textbf{B#1} (#2) #3}

\newcommand\PRD[3]{\textsl{Phys. Rev.} \textbf{D#1} (#2) #3}

\newcommand\PLB[3]{\textsl{Phys. Lett.} \textbf{#1B} (#2) #3}

\newcommand\AoP[3]{\textsl{Ann. of Phys.} \textbf{#1} (#2) #3}

\newcommand\IJMPA[3]{\textsl{Int. J. Mod. Phys.} \textbf{A#1} (#2) #3}

\newcommand\MPLA[3]{\textsl{Mod. Phys. Lett.} \textbf{A#1} (#2) #3}

\begin{document}

\title*{$\bf{f(R)}$-Gravity: ``Einstein Frame'' Lagrangian Formulation, Non-Standard 
Black Holes and QCD-like Confinement/Deconfinement}
\titlerunning{$f(R)$-Gravity, Non-Standard Black Holes and Confinement/Deconfinement} 
\author{Eduardo Guendelman, Alexander Kaganovich, Emil Nissimov, Svetlana Pacheva}
\institute{Eduardo Guendelman and Alexander Kaganovich 
\at Department of Physics, Ben-Gurion University of the Negev, Beer-Sheva, Israel,\\
\email{guendel@bgu.ac.il}, \email{alexk@bgu.ac.il}
\and Emil Nissimov and Svetlana Pacheva 
\at Institute for Nuclear Research and Nuclear Energy, Bulgarian Academy of Sciences, 
Sofia, Bulgaria,
\email{nissimov@inrne.bas.bg}, \email{svetlana@inrne.bas.bg}}
\maketitle


\abstract{We consider $f(R) = R + R^2$ gravity interacting with a dilaton
and a special non-standard form of nonlinear electrodynamics containing a
square-root of ordinary Maxwell Lagrangian. In flat spacetime the latter
arises due to a spontaneous breakdown of scale symmetry and produces
an effective charge-confining potential. In the $R + R^2$ gravity case, 
upon deriving the explicit form of the equivalent {\em local} ``Einstein frame'' 
Lagrangian action, we find several physically relevant features due to the
combined effect of the gauge field and gravity nonlinearities such as:
appearance of dynamical effective gauge couplings and
{\em confinement-deconfinement transition effect} as functions of the
dilaton vacuum expectation value; 
new mechanism for dynamical generation of cosmological constant;
deriving non-standard black hole solutions carrying additional constant vacuum radial 
electric field and with non-asymptotically flat ``hedge-hog''-type spacetime asymptotics.
}

\section{Introduction}
\label{sec:1}
$f(R)$-gravity models (where $f(R)$ is a nonlinear function of the scalar 
curvature $R$ and, possibly, of other higher-order invariants of the Riemann curvature
tensor $R^\k_{\l\m\n}$) are attracting a lot of interest as possible candidates to 
cure problems in the standard cosmological models related to dark matter and dark
energy. For a recent review of $f(R)$-gravity see \textsl{e.g.}
\ct{f(R)-grav} and references therein \foot{The first $R^2$-model (within
the second order formalism), which was proposed as the first inflationary model, 
appeared in Ref.\ct{starobinski}.}.

In the present contribution we consider $f(R)$-gravity coupled to scalar dilaton $\phi$
and most notably -- to a {\em non-standard nonlinear gauge field system containing} 
$\sqrt{-F^2}$ (square-root of standard Maxwell kinetic term; see Refs.\ct{GG-1,GG-2,GG-3}),
which is known to produce confining effective potential among quantized
charged fermions in flat spacetime \ct{GG-2}.

We describe in some detail the explicit derivation of the effective Lagrangian
governing the $f(R)$-gravity dynamics in the so called ``Einstein frame''. 
The latter means that in terms of an appropriate {\em conformal rescaling}
of the original spacetime metric $g_{\m\n} \to h_{\m\n}=f^{\pr}_R\, g_{\m\n}$
(where $f^{\pr}_R = df/dR$) the pertinent gravity part of the effective action assumes 
the standard form of Einstein-Hilbert action ($\sim R(h)$).

Our main goal is to derive a {\em local} ``Einstein frame'' effective Lagrangian 
{\em for the matter fields} as well -- this is explicitly done for ``$R+R^2$-gravity''.

Namely, in the special case of $f(R)=R+\alpha R^2$ the passage to the 
``Einstein frame'' entails non-trivial modifications in the effective matter 
Lagrangian, which {\em in combination with the special ``square-root'' gauge field 
nonlinearity} triggers various physically interesting effects:
\begin{itemize}
\item
(i) appearance of dynamical effective gauge couplings and
{\em confinement-deconfinement transition effect} as functions of the
dilaton vacuum expectation value ( v.e.v.);
\item
(ii) new mechanism for dynamical generation of cosmological constant;
\item
(iii) non-standard black hole solutions carrying a constant vacuum radial electric field 
(such electric fields do not exist in ordinary Maxwell electrodynamics)
and exhibiting  non-asymptotically flat ``hedgehog''-type \ct{hedgehog}
spacetime asymptotics;
\item
(iv) the above non-standard black holes are shown to obey the first law of black hole 
thermodynamics;
\item
(v) obtaining new ``tubelike universe'' solutions of Levi-Civita-Bertotti-Robinson type
$\cM_2 \times S^2$ \ct{LC-BR}.
\end{itemize}

In addition, as shown in Ref.\ct{hide-confine} coupling of the gravity/nonlinear gauge 
field system to {\em lightlike} branes produces ``charge-''hiding'' and 
charge-confining ``thin-shell'' wormhole solutions displaying QCD-like confinement.

The main motivation for including the nonlinear gauge field term $\sqrt{-F^2}$
comes from the works \ct{tHooft} of G. `t Hooft, who has shown that in any effective 
quantum gauge theory, which is able to describe linear confinement phenomena, 
the energy density of electrostatic field configurations should be a linear function
of the electric displacement field in the infrared region (the latter appearing as an
``infrared counterterm'').

The simplest way to realize `t Hooft's  ideas in flat spacetime has been
worked out in Refs.\ct{GG-1,GG-2,GG-3} where the following nonlinear modification of
Maxwell action has been proposed:
\br
S = \int d^4 x \; L(F^2) \quad ,\quad
L(F^2) = -\frac{1}{4} F^2 - \frac{f_0}{2} \sqrt{-F^2} \; ,
\lab{GG-flat} \\
F^2 \equiv F_{\m\n} F^{\m\n} \quad ,\quad 
F_{\m\n} = \pa_\m A_\n - \pa_\n A_\m  \; .
\nonu
\er
The square root of the Maxwell kinetic term naturally arises as a result of 
{\em spontaneous breakdown of scale symmetry} of 
the original scale-invariant Maxwell action with $f_0$ appearing as an integration 
constant responsible for the latter spontaneous breakdown.

For static field configurations the model \rf{GG-flat} yields an electric displacement
field $\vec{D} = \vec{E} - \frac{f_0}{\sqrt{2}}\frac{\vec{E}}{|\vec{E}|}$ and 
the corresponding  energy density turns out to be 
$\h \vec{E}^2 = \h |\vec{D}|^2 + \frac{f_0}{\sqrt{2}} |\vec{D}|+\frac{1}{4} f_0^2$, 
so that it indeed contains a term linear w.r.t. $|\vec{D}|$ as predicted by
the phenomenological theory of `t Hooft.

The {\em non-standard nonlinear} gauge field system \rf{GG-flat} produces
in flat spacetime \ct{GG-2}, when coupled to quantized fermions, a confining 
effective potential $V(r) = - \frac{\b}{r} + \g r$ (Coulomb plus linear one with 
$\g \sim f_0$) which is of the form of the well-known ``Cornell'' potential 
\ct{cornell-potential} in the phenomenological description of quarkonium systems in QCD. 

\section{$f(R)$-Gravity in the ``Einstein Frame''}
\label{sec:2}
Consider $f(R)= R + \a R^2 + \ldots$ gravity (possibly
with a bare cosmological constant $\L_0$) coupled to a dilaton $\phi$ and
a nonlinear gauge field system containing $\sqrt{-F^2}$:
\br
S = \int d^4 x \sqrt{-g} \Bigl\lb \frac{1}{16\pi} 
\Bigl( f\bigl(R(g,\G)\bigr) - 2\L_0 \Bigr) + L(F^2(g)) + L_D (\phi,g) \Bigr\rb \; ,
\lab{f-gravity+GG+D} \\
L(F^2(g)) = - \frac{1}{4e^2} F^2(g) - \frac{f_0}{2} \sqrt{- F^2(g)} \; ,
\lab{GG-g} \\
F^2(g) \equiv F_{\k\l} F_{\m\n} g^{\k\m} g^{\l\n} \;\; ,\;\;
F_{\m\n} = \pa_\m A_\n - \pa_\n A_\m \;
\lab{F2-g} \\
L_D (\phi,g) = -\h g^{\m\n}\pa_\m \phi \pa_\n \phi - V(\phi) \; .
\lab{L-dilaton}
\er
where $R(g,\G) = R_{\m\n}(\G) g^{\m\n}$ and $R_{\m\n}(\G)$ is the Ricci curvature 
in the first order (Palatini) formalism, \textsl{i.e.}, the spacetime metric 
$g_{\m\n}$ and the affine connection $\G^\m_{\n\l}$ are \textsl{a priori} 
independent variables.

The equations of motion resulting from the action \rf{f-gravity+GG+D} read:
\br
R_{\m\n}(\G) = \frac{1}{f^{\pr}_R}\llb 8\pi T_{\m\n} + 
\h f\bigl(R(g,\G)\bigr) g_{\m\n}\rrb \; ,
\lab{g-eqs} \\
f^{\pr}_R \equiv \frac{df(R)}{dR} \quad ,\quad 
\nabla_\l \(\sqrt{-g} f^{\pr}_R g^{\m\n}\)  = 0 \; ,
\lab{gamma-eqs} \\
\pa_\n \Bigl(\sqrt{-g} \Bigl\lb 1/e^2 - \frac{f_0}{\sqrt{-F^2(g)}}
\Bigr\rb F_{\k\l} g^{\m\k} g^{\n\l}\Bigr)=0 \; .
\lab{GG-eqs-R2}
\er
The total energy-momentum tensor is given by:
\br
T_{\m\n} = 
\Bigl\lb L(F^2(g))+L_D (\phi,g)-\frac{1}{8\pi}\L_0\Bigr\rb g_{\m\n} 
\lab{T-total} \\
+ \Bigl(1/e^2 - \frac{f_0}{\sqrt{-F^2(g)}}\Bigr) 
F_{\m\k} F_{\n\l} g^{\k\l} + \pa_\m \phi \pa_\n \phi \;.
\nonu
\er

Eq.\rf{gamma-eqs} leads to the relation $\nabla_\l \( f^{\pr}_R g_{\m\n}\)=0$
and thus it implies transition to the physical ``Einstein frame'' metrics 
$h_{\m\n}$ via conformal rescaling of the original metric $g_{\m\n}$ 
\ct{olmo-etal}:
\be
g_{\m\n} = \frac{1}{f^{\pr}_R} h_{\m\n} \quad ,\quad
\G^\m_{\n\l} = \h h^{\m\k} \(\pa_\n h_{\l\k} + \pa_\l h_{\n\k} - \pa_\k h_{\n\l}\)
\; .
\lab{einstein-frame}
\ee
Using \rf{einstein-frame} the $f(R)$-gravity equations of motion \rf{g-eqs} can be
rewritten in the form of {\em standard} Einstein equations:
\be
R_{\m\n} = 8\pi \({T_{\rm eff}}_{\m\n} - \h g_{\m\n} {T_{\rm eff}}\)
\lab{einstein-eff-eqs}
\ee
where ${T_{\rm eff}} = g^{\m\n} {T_{\rm eff}}_{\m\n}$
and with effective energy-momentum tensor ${T_{\rm eff}}_{\m\n}$ of the following form:
\be
{T_{\rm eff}}_{\m\n} = \frac{1}{f^{\pr}_R} \Bigl\lb T_{\m\n} -
\frac{1}{4}g_{\m\n} T\Bigr\rb - \frac{1}{32\pi} g_{\m\n} R(T) \; .
\lab{T-eff}
\ee
Here $T \equiv g^{\m\n}T_{\m\n}$, $R(T)$ is the original scalar curvature
determined as function of $T$ from the trace of Eq.\rf{g-eqs}:
\be
8\pi T = R f^{\pr}_R - 2 f(R) \; ,
\lab{T-R-eq}
\ee
and everywhere in \rf{einstein-eff-eqs}--\rf{T-R-eq} $g_{\m\n}$ and $\G^\m_{\n\l}$ are
understood as functions of the ``Einstein frame'' metric $h_{\m\n}$ \rf{einstein-frame}.

\section{Einstein-Frame Effective Action}
\label{sec:3}
We are now looking for an effective action 
$S_{\rm eff} = \int d^4 x \sqrt{-h} \Bigl\lb \frac{1}{16\pi} R(h) + L_{\rm eff}\Bigr\rb$,
where $R(h)$ is the standard Ricci scalar of the ``Einstein frame'' metric $h_{\m\n}$ 
and $L_{\rm eff} \equiv L_{\rm eff} (h_{\m\n}, A_\m,\phi)$ is a {\em local function} of 
the corresponding (matter) fields and of their derivatives, such that it produces in 
the standard way the original $f(R)$-gravity equations of motion \rf{g-eqs} 
(or equivalently \rf{einstein-eff-eqs}--\rf{T-R-eq}). $L_{\rm eff}$ will
also include an {\em effective} cosmological constant term irrespective of
the presence or absence of a bare cosmological constant $\L_0$ in the
original $f(R)$-gravity action \rf{f-gravity+GG+D}.

$L_{\rm eff}$ must obey the following relation to the ``Einstein frame''
effective energy-momentum tensor \rf{T-eff}:
\be
{T_{\rm eff}}_{\m\n} = h_{\m\n} L_{\rm eff} - 2 \partder{L_{\rm eff}}{h^{\m\n}} \; .
\lab{T-L-eff}
\ee
Henceforth we will explicitly consider the simplest nonlinear 
$f(R)$-gravity case: $f(R) = R + \a R^2$ (so that $f^{\pr}_R = 1 + 2\a R$).

The generic form of the matter Lagrangian reads:
\be
L_m = L^{(0)} + L^{(1)}(g) + L^{(2)}(g) \; ,
\lab{L-hom}
\ee
where the superscripts indicate homogeneity degree w.r.t. $g^{\m\n}$.
Solving relation \rf{T-L-eff} by taking into account the conformal rescaling
of $g_{\m\n}$ \rf{einstein-frame}  and the homogeneity relation \rf{L-hom} we find the 
following {\em local} effective ``Einstein frame'' matter Lagrangian:
\be
L_{\rm eff} = \frac{1}{1-64\pi\a L^{(0)}} \Bigl\lb L^{(0)} +
L^{(1)}(h) \bigl( 1 + 16\pi\a L^{(1)}(h)\bigr)\Bigr\rb + L^{(2)}(h) \; ,
\lab{L-eff-generic}
\ee
where now the superscripts indicate homogeneity degree w.r.t. $h^{\m\n}$.

Explicitly, in the case of $R+R^2$-gravity/nonlinear-gauge-field/dilaton system
\rf{f-gravity+GG+D}--\rf{L-dilaton} we have
(using shortcut notations $F^2(h) \equiv F_{\k\l} F_{\m\n} h^{\k\m} h^{\l\n}$ and 
$X(\phi,h) \equiv -\h h^{\m\n}\pa_\m \phi \pa_\n \phi$):
\br
L_{\rm eff} = - \frac{1}{4 e_{\rm eff}^2 (\phi)} F^2(h) 
- \h f_{\rm eff} (\phi) \sqrt{- F^2(h)} 
\nonu \\
+ \frac{X(\phi,h)\bigl(1+16\pi\a X(\phi,h)\bigr) - V(\phi) -\L_0/8\pi
}{1+8\a\( 8\pi V(\phi)+\L_0\)}
\lab{L-eff-h}
\er
with the dynamically generated dilaton $\phi$-dependent couplings:
\br
\frac{1}{e_{\rm eff}^2 (\phi)} = \frac{1}{e^2} + 
\frac{16\pi\a f_0^2}{1 + 8\a \(8\pi V(\phi) + \L_0 \)} \; ,
\lab{e-eff} \\
f_{\rm eff}(\phi)=f_0 \frac{1+32\pi\a X(\phi,h)}{1 + 8\a\(8\pi V(\phi)+\L_0\)}
\; .
\lab{f-eff}
\er

Here is an important observation about the effective action:
\be
S_{\rm eff} = \int d^4 x \sqrt{-h} \Bigl\lb \frac{R(h)}{16\pi} 
+ L_{\rm eff} (h,A,\phi)\Bigr\rb \; .
\lab{einstein-frame-action}
\ee
Even if ordinary kinetic Maxwell term
$-\frac{1}{4}F^2$ is absent in the original system ($e^2 \to \infty$ in \rf{GG-g}),
such term is nevertheless {\em dynamically generated} in the ``Einstein frame'' action 
\rf{L-eff-h}--\rf{einstein-frame-action} -- an explicit manifestation of the
{\em combined effect} of gravitational and gauge field nonlinearities 
($\a R^2$ and $-\frac{f_0}{2}\sqrt{-F^2}$):
\be
S_{\rm maxwell} =
-4\pi\a f_0^2 \int d^4 x \sqrt{-h} \frac{F_{\k\l} F_{\m\n} h^{\k\m} h^{\l\n}}{
1+8\a\(8\pi V(\phi)+\L_0\)} \; .
\lab{dynamical-maxwell}
\ee

\section{Confinement/Deconfinement Phases}
\label{sec:4}
In what follows we consider constant dilaton $\phi$ extremizing the effective
Lagrangian \rf{L-eff-h} (\textsl{i.e.}, the dilaton kinetic term $X(\phi,h)$ will
be ignored in the sequel):
\br
L_{\rm eff} =
- \frac{1}{4 e_{\rm eff}^2 (\phi)} F^2(h) - \h f_{\rm eff} (\phi) \sqrt{-F^2(h)}
- V_{\rm eff}(\phi) \; ,
\lab{L-eff-0} \\
V_{\rm eff}(\phi) = \frac{V(\phi) + \frac{\L_0}{8\pi}}{1+8\a\(8\pi V(\phi)+\L_0\)}
\; ,
\lab{V-eff-1} \\
f_{\rm eff} (\phi) = \frac{f_0}{1+8\a\(8\pi V(\phi)+\L_0\)} \; ,
\lab{f-eff-1} \\
\frac{1}{e_{\rm eff}^2 (\phi)} = 
\frac{1}{e^2} + \frac{16\pi\a f_0^2}{1 + 8\a \(8\pi V(\phi) + \L_0 \)} \; .
\lab{e-eff-1}
\er

Here we uncover the following important property: 
{\em the dynamical couplings and the effective potential are extremized 
simultaneously}, which is an explicit realization of the so called 
``least coupling principle'' of Damour-Polyakov \ct{damour-polyakov}:
\be
\partder{f_{\rm eff}}{\phi} = - 64\pi\a f_0 \partder{V_{\rm eff}}{\phi}
\;\; ,\;\; \partder{}{\phi}\frac{1}{e_{\rm eff}^2} =
-(32\pi\a f_0)^2 \partder{V_{\rm eff}}{\phi} 
\;\; \to \partder{L_{\rm eff}}{\phi} \sim \partder{V_{\rm eff}}{\phi} \; .
\lab{f-e-extremize}
\ee
Therefore, at the extremum of $L_{\rm eff}$ \rf{L-eff-0} $\phi$ must satisfy:
\be
\partder{V_{\rm eff}}{\phi} = 
\frac{V^{\pr}(\phi)}{\llb 1+8\a\(\k^2 V(\phi)+\L_0\)\rrb^2} = 0 \; .
\lab{V-extremum}
\ee

There are two generic cases:

(A) {\em Confining phase}: Eq.\rf{V-extremum} is satisfied for some 
finite value $\phi_0$ extremizing the original potential $V(\phi)$: 
$V^{\pr}(\phi_0) = 0$.

(B) {\em Deconfinement phase}: For polynomial or exponentially growing original 
potential $V(\phi)$, so that $V(\phi) \to \infty$ when $\phi \to \infty$, we have:
\be
\partder{V_{\rm eff}}{\phi} \to 0 \quad ,\quad 
V_{\rm eff} (\phi) \to \frac{1}{64\pi\a} = {\rm const} \quad {\rm when} \;\;
\phi \to \infty \; ,
\lab{flat-region}
\ee
\textsl{i.e.}, for sufficiently large values of $\phi$ we find a {\em ``flat region''}
in the effective potential $V_{\rm eff}$. This ``flat region'' triggers a 
{\em transition from confining to deconfinement dynamics}.

Namely, in the confining phase (A) (generic minimum $\phi_0$ of the effective dilaton
potential)
we have shown in \ct{grav-cornell} that the following {\em confining potential} 
(linear w.r.t. $r$) acts on charged test point-particles:
\be
\frac{\sqrt{2}\cE |q_0|}{m_0^2} e_{\rm eff}(\phi_0) f_{\rm eff}(\phi_0)\, r \; ,
\lab{lin-conf}
\ee
where $\cE, m_0, q_0$ are energy, mass and charge of the test particle.

In the deconfinement phase (B) (``flat-region'' of the effective dilaton potential)
we have:
\be
f_{\rm eff} \to 0 \quad ,\quad e^2_{\rm eff} \to e^2
\lab{deconfine}
\ee
and the effective gauge field Lagrangian \rf{L-eff-0} reduces to the ordinary
\textsl{non-confining} one (the ``square-root'' term $\sqrt{-F^2}$ vanishes):
\be
L^{(0)}_{\rm eff} = -\frac{1}{4e^2} F^2(h) - \frac{1}{64\pi\a}
\lab{L-eff-h-0}
\ee
with an {\em induced} cosmological constant $\L_{\rm eff} = 1/8\a$, which is
{\em completely independent} of the bare cosmological constant $\L_0$.

\section{Non-Standard Black Holes and New ``Tubelike'' Solutions}
\label{sec:5}

From the effective Einstein-frame action \rf{einstein-frame-action} with 
$L_{\rm eff}$ as in \rf{L-eff-0} we find 
\textsl{non-standard} Reissner-Nordstr{\"o}m-(anti-)de-Sitter-type black
hole solutions in the confining phase ($\phi_0$ -- generic minimum of the effective 
dilaton potential \rf{V-eff-1}; $e_{\rm eff}(\phi))$, $f_{\rm eff}(\phi)$ as in
\rf{f-eff-1}--\rf{e-eff-1}):
\br
ds^2 = - A(r) dt^2 + \frac{dr^2}{A(r)} + r^2 \bigl(d\th^2 + \sin^2 \th d\vp^2\bigr)
\; ,
\lab{spherical-static} \\
A(r) = 1 - \sqrt{8\pi}|Q|f_{\rm eff}(\phi_0) e_{\rm eff}(\phi_0) 
- \frac{2m}{r} + \frac{Q^2}{r^2} - \frac{\L_{\mathrm{eff}}(\phi_0)}{3} r^2 \; ,
\lab{CC-eff}
\er
with {\em dynamically generated} cosmological constant:
\be
\L_{\rm eff}(\phi_0) = \frac{\L_0 +8\pi V(\phi_0)+2\pi e^2 f^2_0}{
1+8\a\(\L_0 +8\pi V(\phi_0)+2\pi e^2 f^2_0\)} \; .
\lab{h-CC-eff}
\ee

The black hole's static spherically symmetric electric field contains
apart from the Coulomb term an {\em additional constant radial ``vacuum'' piece}
responsible for confinement (let us stress again that constant vacuum radial electric 
fields do not exist in ordinary Maxwell electrodynamics):
\br
|F_{0r}| = |\vec{E}_{\rm vac}| +
\frac{|Q|}{\sqrt{4\pi}\, r^2} \Bigl(\frac{1}{e^2} + 
\frac{16\pi\a f_0^2}{1 + 8\a\(8\pi V(\phi_0) + \L_0 \)}\Bigr)^{-\h}
\lab{cornell-sol} \\
|\vec{E}_{\rm vac}| \equiv 
\Bigl(\frac{1}{e^2} + 
\frac{16\pi\a f_0^2}{1 + 8\a\(8\pi V(\phi_0) + \L_0 \)}\Bigr)^{-1}
\frac{f_0/\sqrt{2}}{1+8\a\(8\pi V(\phi_0)+\L_0\)} \; .
\lab{vacuum-radial}
\er
For the special value of $\phi_0$ where $\L_{\rm eff}(\phi_0)=0$ we obtain
Reissner-Nordstr{\"o}m-type black hole with a ``hedgehog'' \ct{hedgehog}
{\em non-flat-spacetime} asymptotics: \\
$A(r) \to 1 -\sqrt{8\pi}|Q|f_{\rm eff}(\phi_0) e_{\rm eff}(\phi_0) \neq 1 ~$ 
for $~ r\to\infty$.

Further we obtain Levi-Civitta-Bertotti-Robinson (LCBR) \ct{LC-BR} type ``tubelike''
spacetime solutions with geometries $\cM_2 \times S^2$ ($\cM_2$ -- 2-dimensional manifold)
with metric of the form:
\be
ds^2 = - A(\eta) dt^2 + \frac{d\eta^2}{A(\eta)} 
+ r_0^2 \bigl(d\th^2 + \sin^2 \th d\vp^2\bigr) \;\; ,\;\;  
-\infty < \eta <\infty \; ,
\lab{gen-BR-metric}
\ee
and constant vacuum ``radial'' electric field $|F_{0\eta}| = |\vec{E}_{\rm vac}|$,
where the size of the $S^2$-factor is given by (using short-hand 
$\L(\phi_0)\equiv 8\pi V(\phi_0) + \L_0$):
\be
\frac{1}{r_0^2} = \frac{4\pi}{1+8\a\L(\phi_0)}\Bigl\lb
\Bigl(1+8\a\(\L(\phi_0)+2\pi f_0^2\)\Bigr) \vec{E}_{\rm vac}^2 +
\frac{1}{4\pi}\L(\phi_0)\Bigr\rb \; .
\lab{r0-eq}
\ee

There are three distinct solutions for LBCR \rf{gen-BR-metric}
where $\cM_2 = AdS_2, Rind_2, dS_2$
(2-dimensional anti-de Sitter, Rindler and de Sitter spaces, respectively):

(i) LBCR type solution $AdS_2 \times S^2$ for strong $|\vec{E}_{\rm vac}|$:
\be
A(\eta) = 4\pi K(\vec{E}_{\rm vac}) \eta^2 \quad ,\quad K(\vec{E}_{\rm vac}) >0 \; ,
\lab{AdS2}
\ee
in the metric \rf{gen-BR-metric}, $\eta$ being the Poincare patch
space-like coordinate of $AdS_2$, and
\be
K(\vec{E}_{\rm vac}) \equiv 
\Bigl(1+8\a\(\L(\phi_0)+2\pi f_0^2\)\Bigr) \vec{E}_{\rm vac}^2 
- \sqrt{2}f_0 |\vec{E}_{\rm vac}| - \frac{1}{4\pi}\L(\phi_0) \; .
\lab{K-def}
\ee

(ii) LBCR type solution $Rind_2 \times S^2$ when $K(\vec{E}_{\rm vac})=0$:
\be
A(\eta) = \eta \;\; \mathrm{for}\;\; 0 < \eta < \infty \quad \mathrm{or} \quad
A(\eta) = - \eta \;\; \mathrm{for}\;\; -\infty <\eta < 0 
\lab{Rindler2}
\ee

(iii) LBCR type solution $dS_2 \times S^2$ for weak $|\vec{E}_{\rm vac}|$:
\be
A(\eta) = 1 - 4\pi |K(\vec{E}_{\rm vac})|\,\eta^2 \quad ,\quad 
K(\vec{E}_{\rm vac}) <0 \; .
\lab{dS2}
\ee

\section{Thermodynamics of Non-Standard Black Holes}
\label{sec:6}
Consider static spherically symmetric metric 
$ds^2 = - A(r) dt^2 + \frac{dr^2}{A(r)} + r^2 \bigl(d\th^2 + \sin^2 \th d\vp^2\bigr)$
with Schwarzschild-type horizon $r_0$, \textsl{i.e.}, 
$A(r_0)=0\; ,\; \pa_r A\!\!\bv_{r_0} >0$ and with $A(r)$ of the general 
``non-standard'' form:
\be
A(r) = 1 - c(Q_i) - 2m/r + A_1 (r;Q_i) \; ,
\lab{spherical-static-generic}
\ee
where $Q_i$ are the rest of the black hole parameters apart from the mass $m$, 
and $c(Q_i)$ is generically a non-zero constant (as in \rf{CC-eff}) responsible 
for a ``hedgehog'' \ct{hedgehog} {\em non-flat spacetime asymptotics}.

The so called {\em surface gravity} $\k$ proportional to Hawking temperature 
$T_h$ is given by $\k = 2\pi T_h = \h \pa_r A\!\!\bv_{r_0}$ (cf.,
\textsl{e.g.}, \ct{textbooks}). 

One can straightforwardly derive the first law of black hole thermodynamics
for the non-standard black hole solutions with \rf{spherical-static-generic}:
\be
\d m = \frac{1}{8\pi}\k \d A_H + {\wti\P}_i \d Q_i \; ,\; A_H = 4\pi r_0^2
\; ,\; {\wti\P}_i =
\frac{r_0}{2} \partder{}{Q_i} \Bigl( A_1 (r_0;Q_i) - c(Q_i)\Bigr) \; .
\lab{first-law}
\ee
In the special case of non-standard Reissner-Nordstr{\"o}m-(anti-)de-Sitter 
type black holes \rf{spherical-static}--\rf{h-CC-eff} with parameters $(m,Q)$
the conjugate potential in \rf{first-law}:
\be
{\wti\P} = \frac{Q}{r_0} - 
\sqrt{2\pi} f_{\rm eff} (\phi_0) e_{\rm eff} (\phi_0) r_0  
\equiv \frac{\sqrt{4\pi}}{e_{\rm eff}^2 (\phi_0)} A_0 \bv_{r=r_0}
\lab{conjugate-potential}
\ee
(with $e_{\rm eff} (\phi_0)$ and $f_{\rm eff} (\phi_0)$ as in
\rf{e-eff}--\rf{f-eff}) is (up to a dilaton v.e.v.-dependent factor) the electric 
field potential $A_0$ ($F_{0r} = - \pa_r A_0$) of the nonlinear gauge system on
the horizon.

\section{Conclusions}
\label{sec:7}
In the present contribution we have uncovered a non-trivial interplay between a 
special gauge field non-linearity and $f(R)$-gravity. On one hand, the inclusion of the 
non-standard nonlinear ``square-root'' Maxwell term $\sqrt{-F^2}$ is the explicit 
realization of the old ``classic'' idea of `t Hooft \ct{tHooft} about the nature of 
low-energy confinement dynamics. On the other hand, coupling of the nonlinear gauge 
theory containing $\sqrt{-F^2}$ to $f(R)=R + \a R^2$ gravity plus scalar dilaton
leads to a variety of remarkable effects:

\begin{itemize}
\item
Dynamical effective gauge couplings and dynamical induced cosmological
constant -- functions of dilaton v.e.v..
\item
New non-standard black hole solutions of Reissner-Nordstr{\"o}m-({\em anti}-)de-Sitter 
type carrying an additional constant vacuum radial electric field, in particular, 
non-standard Reissner-Nordstr{\"o}m type black holes with asymptotically non-flat 
``hedgehog'' behaviour.
\item
``Cornell''-type {\em confining} effective potential in charged test particle 
dynamics.
\item
Cumulative simultaneous effect of $\sqrt{-F^2}$ and $R^2$-terms -- triggering
{\em transition from confining to deconfinement phase}.
Standard Maxwell kinetic term for the gauge field $-F^2$ is 
{\em dynamically generated} even when absent in the original ``bare'' theory.
\end{itemize}
Furthermore, as we have shown in Ref.\ct{hide-confine}:
\begin{itemize}
\item
Coupling to a charged lightlike brane produces a charge-``hiding'' wormhole,
where a genuinely charged matter source is detected as electrically
neutral by an external observer.
\item
Coupling to two oppositely charged lightlike brane sources produces a 
two-``throat'' wormhole displaying a genuine QCD-like charge confinement.
\end{itemize}

\begin{acknowledgement}
We gratefully acknowledge support of our collaboration through the academic exchange 
agreement between the Ben-Gurion University and the Bulgarian Academy of Sciences.
S.P. has received partial support from COST action MP-1210.
\end{acknowledgement}
%
%

\end{document}